\documentclass[aip,cha,reprint,onecolumn,author-year]{revtex4-1}

\usepackage{times}
\usepackage{amssymb}
\usepackage{amsmath}
\usepackage{graphicx}
\usepackage{xcolor}
\usepackage{hyperref}
 \hypersetup{colorlinks=true, linkcolor=blue, breaklinks=true, urlcolor=blue, citecolor=blue}
\usepackage{colortbl}

\begin{document}

\title{Heavy-Tailed Distribution of the Number of Papers within Scientific Journals}
\author{Robin Delabays}
 \homepage{orcid.org/0000-0001-6344-1207}
 \email{robindelabays@ucsb.edu.}
 \affiliation{Center for Control, Dynamical Systems and Computation, UC Santa Barbara, Santa Barbara, CA 93106 USA.}

\author{Melvyn Tyloo}
 \homepage{orcid.org/0000-0003-1761-4095}
 \affiliation{Theoretical Division, Los Alamos National Laboratory, Los Alamos, NM 87545 USA.}

\date{\today}

\begin{abstract}
 Scholarly publications represent at least two benefits for the study of the scientific community as a social group. 
 First, they attest of some form of relation between scientists (collaborations, mentoring, heritage,...), useful to determine and analyze social subgroups. 
 Second, most of them are recorded in large data bases, easily accessible and including a lot of pertinent information, easing the quantitative and qualitative study of the scientific community. 
 Understanding the underlying dynamics driving the creation of knowledge in general, and of scientific publication in particular can contribute to maintaining a high level of research, by identifying good and bad practices in science. 
 In this article, we aim at advancing this understanding by a statistical analysis of publication within peer-reviewed journals. 
 Namely, we show that the distribution of the number of papers published by an author in a given journal is heavy-tailed, but has lighter tail than a power law. 
 Interestingly, we demonstrate (both analytically and numerically) that such distributions match the result of an modified preferential attachment process, where, on top of a Barab\'asi-Albert process, we take finite career span of scientists into account. 
 
 ~
  
 {\bf Keywords:} Heavy-tail, publications, scholarly journals, preferential attachment, cumulative advantage. 
\end{abstract}

\maketitle 

\section{Introduction}

One of the core mechanism in the practice of science is the self examination of a field of research. 
The validation of a scientific result is always collective, in the sense that it has been scrutinized, criticized, and (hopefully) validated by a sufficient number of peers. 
Furthermore, any scientific result is permanently subject to new evaluation and might be replaced by a more accurate work. 
At the level of a community, scientists are then used to criticize the work of colleagues and to have their work criticized by them. 
It is then not surprising that some scientists started to study (and thus somehow critically assess) the scientific community itself~[\cite{deS63}]. 

The quantitative study of the scientific community, sometimes referred to as \emph{Science of Science}~[\cite{deS76,Nar76,For18,van19}], is a key step to unravel the underlying behaviors of its composing agents (authors, journals, institutions, etc.). 
Pioneered by the early works of~\cite{Lot26}, the science of science gained a lot of momentum in the second half of the XXth century, with the creation of the first data bases of scientific publications [\cite{Gar55,deS65,Mer68}]. 
More recently, the scientometric investigations have been significantly eased by the emergence of large online data bases of scientific publications (Web of Science, PubMed, arXiv,...) and the ever increasing computation power of modern computers. 
These improvements allowed to analyze scientometric indicators on a larger scale [\cite{Wan16,Fra17}] and with finer resolution in terms of publication units (considering single articles instead of whole journals [e.g., \cite{Wal12b}]) and time [\cite{Egg00,New01a}]. 
For a clear historical overview of scientometrics, we refer to~\cite{van19}. 

The science of science has the potential to help maintaining the quality of research, and thus a good use of public funding. 
There is nowadays an increasing number of scientific papers~[\cite{deS65,Bor15}], combined with the ubiquitous presence of \emph{predatory journals} which publish the papers they receive, charging publication fees, but without performing the fundamental editorial work that guarantees the papers' quality (e.g., quality and pertinence check, referee process)~[\cite{Boh13,Sor17}]. 
In such a context, distinguishing bad practices from honest work in scientific publishing becomes more and more challenging. 
Understanding the underlying dynamics of scientific publication will be instrumental in this endeavor. 

The fight against predatory publishing has benefited from the effort of many dedicated citizens, whose initiatives have shown their efficiency~[\cite{But13,Gru19}], as well as their limits~[\cite{Bea17}]. 
In regard of the proliferation of predatory journals, the task of identifying all of them unequivocally is overwhelming. 
In such a context, the ability to perform a preliminary data-based sanity check of a given journal would allow to focus the resources on the more problematic venues. 
However, such an approach requires an accurate understanding of the quantitative and qualitative characteristics of scientific journals which is still scarce. 

The quality of a scientist's work is commonly quantified by two different, but related, measures. 
Namely, their number of papers and the number of citations thereof (summarized in the \emph{h}-index~[\cite{Hir05,Siu20}]).
A vast majority of investigations about the scientific publication process is focused on the citation side. 
These analysis mostly aim at describing how the citation network impacts the number of citations a given paper is (and therefore its authors are) likely to receive. 
In particular, evidence suggests that citations follow a \emph{cumulative advantage} or \emph{preferential attachment} process, where the more citations a scientist has, the more likely they are to get new citations~[\cite{deS76}]. 
This process leads to a power law distribution of citations~[\cite{Eom11,Wal12a}] or other heavy-tailed distributions~[\cite{The16}]. 
Indeed, preferential attachment has been proven to lead to heavy-tailed distributions~[\cite{Kra00}], with some refinements to account for the life-time of a paper~[\cite{Par15}].

As early as 1926, Lotka showed that, in the field of chemistry, the number of scientists having published $N$ papers is proportional to $N^{-2}$~[\cite{Lot26}]. 
In other words, he showed that the distribution of the number of papers published by scientists follows a power law. 
Later on, the same analysis has been extended to other fields of science~[e.g., \cite{Gup96,Wag99,Hub01a,Hub01b,Sut01,New03,Bar08,Pal15}] and refined to more elaborate distributions, such as the \emph{power law with cutoff}~[\cite{Saa99,Kre01,Smo17}] or the \emph{stretched exponential} distribution~[\cite{Lah98}]. 
Despite this early start, the number of papers published by a scientist has been less investigated than the number of citations that a paper or a scientist gets.

\begin{figure*}
 \centering
 \includegraphics[width=\textwidth]{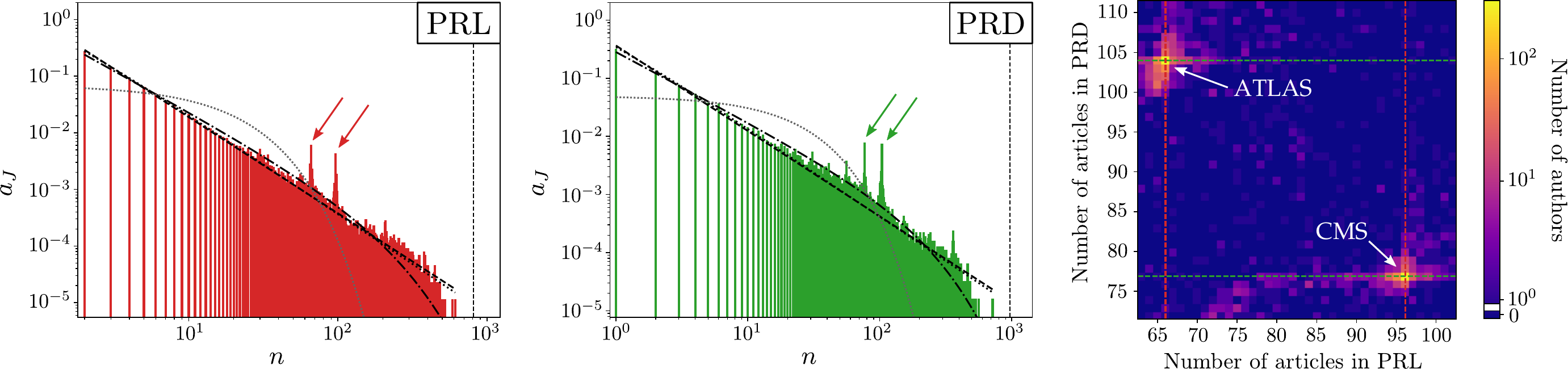}
 \caption{Left and center: histograms of the number of papers $n$ published in \emph{Phys. Rev. Lett.} (PRL) and \emph{Phys. Rev. D} (PRD) among the authors who published in these journals. 
 For each value of $n$, the height of the bar gives the proportion of authors who published $n$ articles in the corresponding journal. 
 Best distribution fits (see Sec.~\ref{sec:fit}) are displayed for an exponential distribution (gray dotted), a power law (dashed black), an power law with cutoff (dash-dotted black), and a Yule-Simon distribution (dotted black).   
 The arrows indicate significant peaks in the number of authors corresponding to the ATLAS and CMS experiments at the CERN. 
 Right: Two-dimensional, color-coded histogram of the number of authors with respect to the number of papers published in PRL (horizontal axis) and PRD (vertical axis). 
}
 \label{fig:prl_prd}
\end{figure*}

With the objective of refining these past analysis, in this article, we focus on the distribution of the number of papers published by scientists within a given peer-reviewed journal. 
The distribution of the number of papers is both easily accessible (through any scientific publication data base) and informative. 
Indeed, various characteristics of the publication dynamics within a journal can be extracted from the aforementioned distribution. 
We illustrate this claim in the striking examples of \emph{Physical Review Letters} and \emph{Physical Review D}, shown in Fig.~\ref{fig:prl_prd}, where the analysis of the distribution emphasizes: (i) an underlying \emph{preferential attachment} dynamics; (ii) the finiteness of the scientific careers; and (iii) the presence of (very) large groups of scientists in the related fields of physics (see caption of Fig.~\ref{fig:prl_prd} for a detailed discussion).

As interestingly pointed out by~\cite{Sek18}, publishing in a peer-reviewed journal (especially in high-impact ones) is more likely if one author of the manuscript already published in the same journal.
Such a process can be interpreted as \emph{preferential attachment}, and an expected outcome of such an observation is a high representation of a few authors in a given journal~[\cite{Kra00}]. 
Furthermore, a scientist whose field of research is well-aligned with a journal topic is likely to publish a large proportion of their work in this journal, leading again to a high representation of a few specialized authors in a given journal. 

\begin{figure*}
 \centering
 \includegraphics[width=\textwidth]{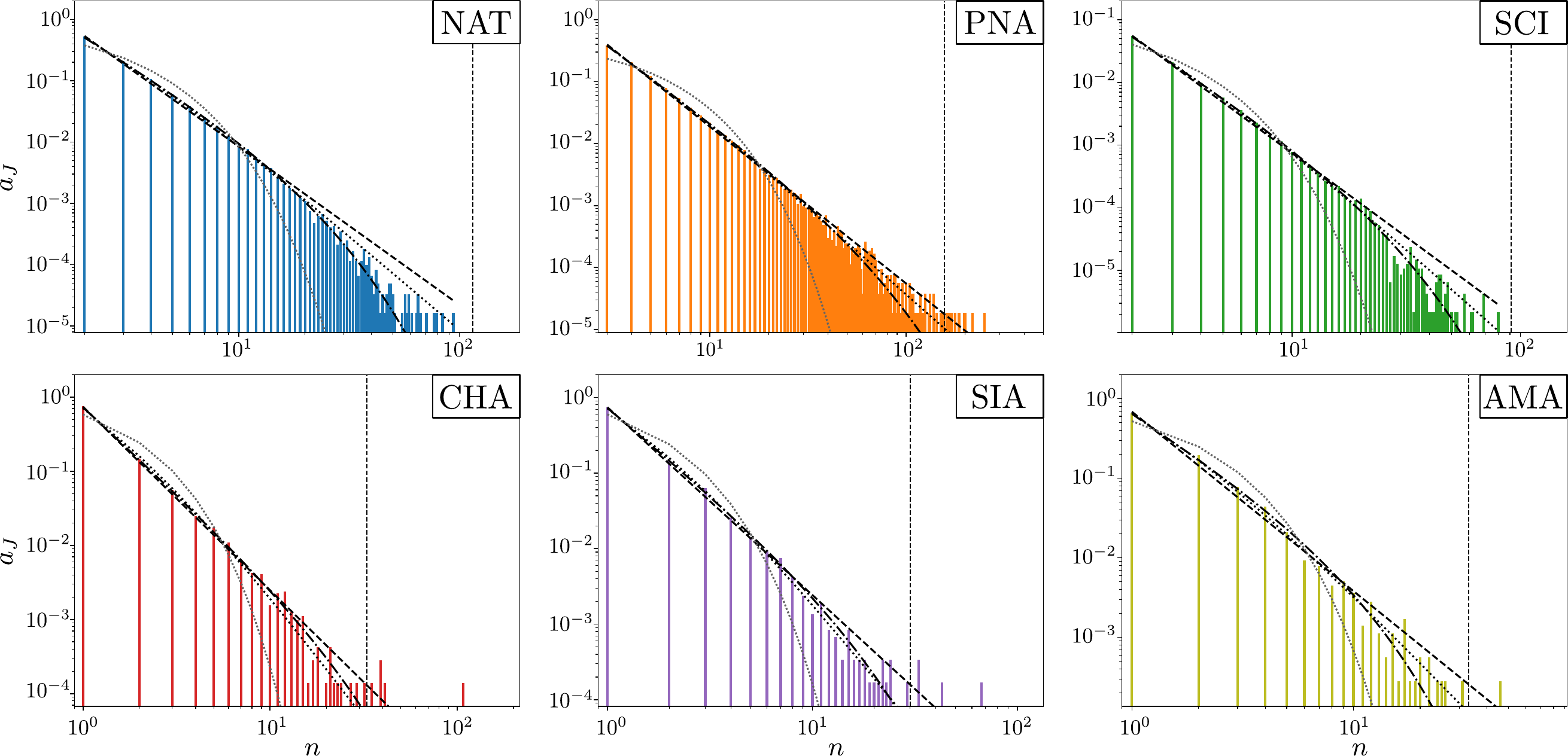}
 \caption{Histograms of the number of papers $n$ published in the six journals indicated in the insets, among the authors who published in these journals (see Table~\ref{tab:journals} for legends). 
 As in Fig.~\ref{fig:prl_prd}, for each value of $n$, the height of the bar gives the proportion of authors who published $n$ articles in the corresponding journal. 
 The gray dotted line is an exponential fit of the data, emphasizing that the distribution is heavy-tailed. 
 We also show the best fit (MLE), discussed in Sec.~\ref{sec:fit}, for a power law distribution (dashed black), power law with cutoff (dash-dotted black), and Yule-Simon distribution (dotted black). 
 The vertical dashed line indicates the theoretical maximal number of papers if the distribution was the fitted power law [see Sec.~\ref{sec:kplayers}). 
 The same plots for the other journals are available in Fig.~\ref{fig:prl_prd} and in the Supplementary Figure~\ref{fig:s1}.} 
 \label{fig:full12}
\end{figure*}

The heavy-tailedness of the distribution of the number of papers is striking in the histograms (see Figs.~\ref{fig:prl_prd} and \ref{fig:full12}). 
Indeed, the tail of the histogram is stronger than the best exponential fit to the data (gray dotted line). 
However, as we show below, the famous \emph{power law} is not a good fit to the data neither, and the actual distribution lies somewhere between an exponential and a power law. 
In addition to our analysis of the distribution, we propose an adaptation of the preferential attachment law that models the evolution of the number of papers of a set of authors, within a journal. 

\section{Empirical and fitted distributions}

\begin{table}
 \centering
 \begin{tabular}{l|l|l}
  Label & Journal name (red. year) & \# authors (red.) \\
  \hline
  \hline NAT & Nature$^*$ (1950) & 63'791 (3'374) \\
  \hline PNA & Proc. Natl. Acad. Sci. USA$^{**}$ (1950) & 55'849 (2'495) \\
  \hline SCI & Science$^*$ (1940) & 48'928 (4'788) \\
  \hline LAN & The Lancet$^*$ (1910) & 33'416 (3'015) \\
  \hline NEM & New England Journal of Medicine$^*$ (1950) & 27'078 (3'842) \\
  \hline PLC & Plant Cell (2000) & 20'649 (4'712) \\
  \hline ACS & J. of the American Chemical Society$^*$ (1930) & 82'223 (5'301) \\
  \hline TAC & IEEE Trans. on Automatic Control (2000) & 8'911 (3'603) \\
  \hline ENE & Energy (2005) & 28'920 (4'491) \\
  \hline CHA & Chaos & 7'409 \\
  \hline SIA & SIAM Journal on Applied Mathematics & 6'106 \\
  \hline AMA & Annals of Mathematics & 3'679 \\
  \hline PRD & Physical Review D & 64'922  \\
  \hline PRL & Physical Review Letters$^*$ & 90'993 
 \end{tabular}
 \caption{Labels, names, and number of authors in the journals considered. 
 In parenthesis is given the reduction year (discussed in Sec.~\ref{sec:kplayers}) and the number of authors up to this year. 
 One (resp. two) asterisk(s) indicate the journals where authors with one (resp. two) paper(s) are discarded.}
 \label{tab:journals}
\end{table}

We consider an arbitrary selection of 14 peer-reviewed journals (Table~\ref{tab:journals}), whose data are available on the Web of Science data base (WoS, \url{www.webofscience.com}). 
The selected journals vary in age (from a few decades to more than a century) but are not too young, in order to have sufficiently many papers available, and all of them are still publishing nowadays. 
Whereas the choice of journals is arbitrary and limited, we tried to cover a diversity of disciplines of the natural sciences and various time spans. 
The limited sample of journals do not allow us to be claim any universality in our results, but we argue that it demonstrate the pertinence of our approach in the quantitative analysis of the scientific publication process. 

We denote by ${\cal J}=\{{\rm NAT},{\rm PNA},...,{\rm PRL}\}$ the set of journals considered (see Table~\ref{tab:journals} for the list of labels).
Within each journal $J\in {\cal J}$, we index authors by an integer $i=1,...,A_J^{\rm tot}$, $A_J^{\rm tot}$ being the number of authors who published in journal $J$. 
Then for each author $i=1,...,A_J^{\rm tot}$, we count the number $n^J_i$ of papers published by author $i$ in journal $J$ up to year $2017$ in the whole WoS data base (meaning from year 1900 or the year of the journal's creation, whichever is the latest). 
This process yields the set of data ${\cal D}_J=\{n^J_i\colon~i=1,...,A_J^{\rm tot}\}$, which is a set of $A_J^{\rm tot}$ integer numbers. 
We restrict our investigation to papers labeled as \emph{``Article''} in the WoS data base, to focus on peer-reviewed papers. 

From the data set ${\cal D}_J$ we can compute the number and proportion of authors who published $n$ papers 
\begin{align}
 A_J(n) &= \#\{i\colon n^J_i = n\}\, , & a_J(n) &= A_J(n)/A_J^{\rm tot}\, ,
\end{align}
and by definition, $\sum_na_J(n) = 1$. 
The proportion $a_J$ is represented in logarithmic scales in Figs.~\ref{fig:prl_prd}, \ref{fig:full12}, and \ref{fig:s1}, each panel corresponding to a different journal. 

~

{\bf Remark.}
{\it Note that we did not take into account the fact the different papers are co-signed by multiple authors. 
Consequently, different papers have different "weights" in the data set. 
This article is mostly interested in the number of papers from the point of view of the authors, it is then adequate to count, for each author, the number of paper they signed, independently of the number of co-authors. 
Refining the analysis and taking into account the number of co-authors on each paper would be the purpose of future work. 

Note also that we do not take into account papers published anonymously, which represent a large number of papers in medicine journals in particular. 

Finally, for some journals, the number of authors is too large to be downloaded from the WoS data base. 
As a consequence, the authors having published only one or two papers in these journals have to be removed from the data (e.g., ${\rm NAT}$, ${\rm PNA}$, or ${\rm SCI}$, indicated by asterisks in Table~\ref{tab:journals}).
}

\subsection{Distribution fitting}\label{sec:fit}
In regard of the apparent heavy-tailedness of the distribution, it is tempting to fit a power law. 
However, as pointed out by~\cite{Cla09}, such fitting should be done with care in order to avoid spurious conclusions~[\cite{Bro19}]. 
We therefore fit three heavy-tailed distributions and assess the goodness-of-fit of our fitting following~\cite{Cla09}, which is encoded in a $p$-value. 
Numerical results are summarized in Table~\ref{tab:fit_gof}.

For each empirical distribution of the number of papers published by an author $i$ in journal $J$, we fit an exponential distribution (gray dotted lines in Figs.~\ref{fig:prl_prd} and \ref{fig:full12}) to emphasize their heavy-tailed behavior. 
The three heavy-tailed distribution that we fit are:
\begin{itemize}
 \item A \emph{power law distribution} (black dashed lines in the figures), 
 \begin{align}\label{eq:pl}
  {P}_{\rm pl}(n_i^J = n;\alpha) &= C_{\alpha} n^{-\alpha}\, ,
 \end{align}
 with $\alpha>1$ and $C_\alpha\in\mathbb{R}$ normalizing the distribution;
 \item A \emph{power law with cutoff} (black dash-dotted lines in the figures), 
 \begin{align}\label{eq:plco}
  {P}_{\rm plc}(n_i^J = n;\beta,\gamma) &= C_{\beta,\gamma} n^{-\beta}e^{-\gamma n}\, ,
 \end{align} 
 with $\beta>1$, $\gamma>0$, and normalizing constant $C_{\beta,\gamma}\in\mathbb{R}$;
 \item A \emph{Yule-Simon distribution} (black dotted lines in the figures), 
 \begin{align}\label{eq:yule}
  {P}_{\rm ys}(n_i^J = n;\rho) &= C_\rho (\rho-1){\rm B}(n,\rho)\, ,
 \end{align}
 with $\rho>0$, $C_\rho\in\mathbb{R}$ is the normalizing constant, and where ${\rm B}(x,y)$ is the \emph{Euler beta function}. 
\end{itemize}

We perform the distribution fitting by optimizing the parameters $\alpha$, $\beta$, $\gamma$, and $\rho$ with a Maximum Likelihood Estimator~[\cite{Cla09}]. 
The curves of the fitted distributions are plotted in Figs.~\ref{fig:prl_prd}, \ref{fig:full12}, and in the supplementary figure \ref{fig:s1}, and the fitted parameters are given in Table~\ref{tab:fit_gof}. 
Other distributions (such as log-normal, L\'evy, Weibull) were tested and discarded because they were far from matching the data. 

\begin{table}
 \centering
 \begin{tabular}{l||c|c||c|c|c||c|c}
  & \multicolumn{2}{c||}{PL} & \multicolumn{3}{c||}{PLwC} & \multicolumn{2}{c}{Y-S} \\
  & $\alpha$ & $p$ [\%] & $\beta$ & $\gamma$ & $p$ [\%] & $\rho$ & $p$ [\%] \\
  \hline
  \hline NAT & $2.58$ & $0.0$ & $2.11$ & $0.07$ & $0.0$ & $3.10$ & $0.0$ \\
  \hline PNA & $2.53$ & $0.0$ & $2.30$ & $0.02$ & $0.0$ & $2.83$ & $0.0$ \\
  \hline SCI & $2.68$ & $0.0$ & $\bf 2.30$ & $\bf 0.06$ & $\bf 16.64$ & $3.28$ & $0.02$ \\ 
  \hline LAN & $2.47$ & $0.0$ & $2.09$ & $0.05$ & $0.18$ & $2.90$ & $0.0$ \\
  \hline NEM & $2.76$ & $0.0$ & $2.36$ & $0.07$ & $0.2$ & $\bf 3.43$ & $\bf 8.82$ \\
  \hline PLC & $2.30$ & $0.0$ & $\bf 1.92$ & $\bf 0.10$ & $\bf 13.42$ & $3.01$ & $0.92$ \\
  \hline ACS & $2.11$ & $0.0$ & $1.95$ & $0.01$ & $0.0$ & $2.32$ & $0.0$ \\
  \hline TAC & $2.08$ & $0.0$ & $1.84$ & $0.04$ & $0.0$ & $2.51$ & $0.02$ \\
  \hline ENE & $2.36$ & $0.0$ & $2.12$ & $0.06$ & $0.12$ & $3.15$ & $0.0$ \\
  \hline CHA & $2.47$ & $0.0$ & $\bf 2.28$ & $\bf 0.05$ & $\bf 80.84$ & $3.43$ & $0.0$ \\
  \hline SIA & $2.49$ & $0.0$ & $2.20$ & $0.08$ & $2.24$ & $\bf 3.49$ & $\bf 9.06$ \\
  \hline AMA & $2.26$ & $0.0$ & $1.72$ & $0.14$ & $0.18$ & $2.95$ & $0.0$ \\
  \hline PRD & $1.49$ & $0.0$ & $1.24$ & $0.005$ & $0.02$ & $1.55$ & $0.0$ \\
  \hline PRL & $1.73$ & $0.0$ & $1.52$ & $0.005$ & $0.12$ & $1.80$ & $0.0$  
 \end{tabular}
 \caption{Fitted parameters and $p$-value of the goodness-of-fit for power law (PL), power law with cutoff (PLwC), and Yule-Simon (Y-S) distributions. 
 No set of data is well-fitted by a power law distribution. 
 However, the power law with cutoff seems to be a good fit for three journals (SCI, PLC, CHA), and the Yule-Simon distribution seems to correctly fit the distribution of NEM and SIA. 
 For the other journals, none of the distributions seem to fit the data appropriately. }
 \label{tab:fit_gof}
\end{table}

\subsection{Goodness-of-fit}
\begin{figure}
 \centering
 \includegraphics[width=.5\textwidth]{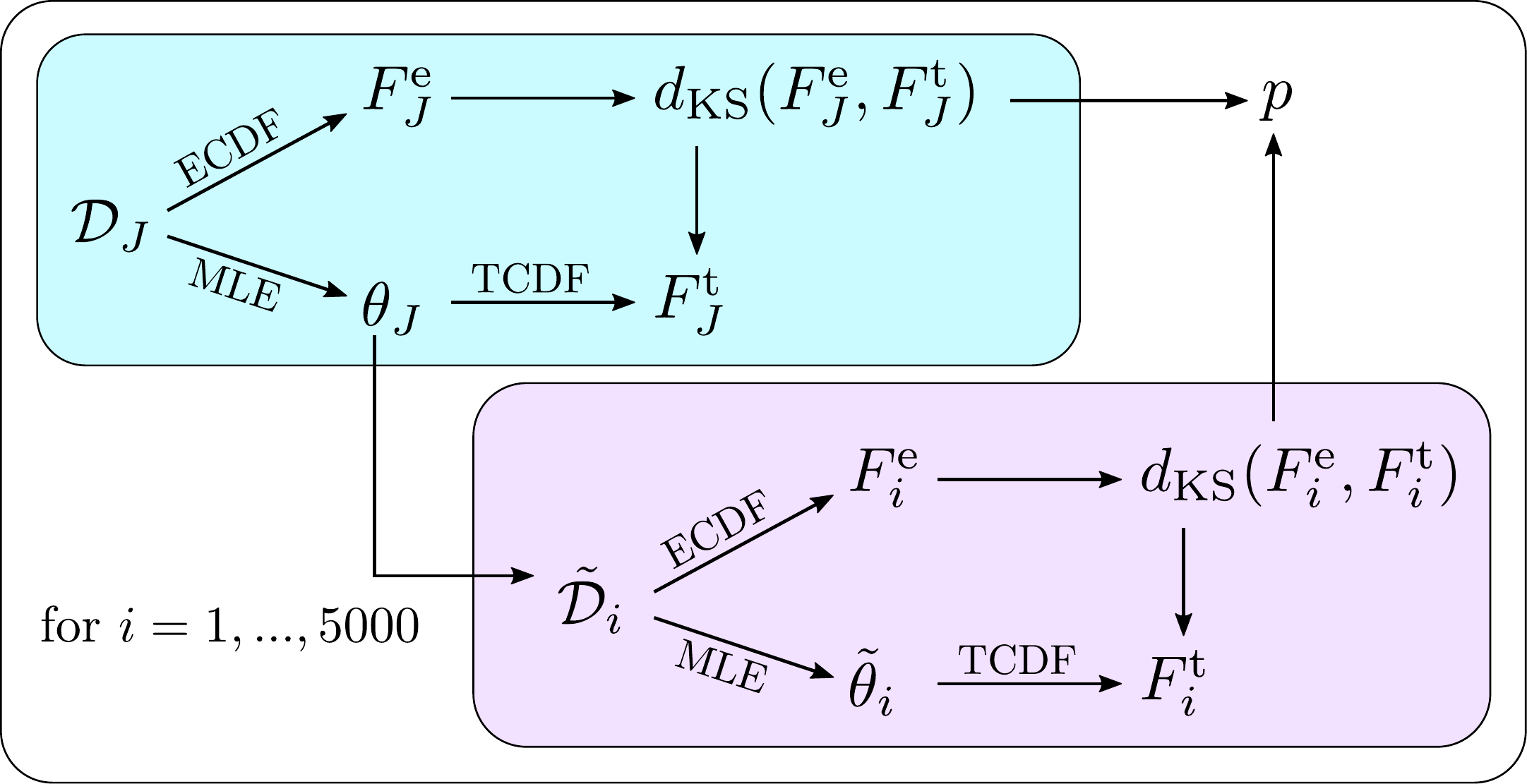}
 \caption{Scheme of the goodness-of-fit computation. 
 For a given journal $J$, the data set ${\cal D}_J$ is fitted with a distribution whose parameters are $\theta_J$, and we compute the Kolmogorov-Smirnov (KS) distance between its empirical and theoretical cumulative distribution functions. 
 Then, based on the parameters $\theta_J$, we generate $5000$ synthetic data sets $\tilde{\cal D}_i$ for $i=1,...,5000$, on which we repeat the same process. 
 Finally, the $p$ value is the proportion of sythetic data sets whose empirical and theoretical cumulative distribution functions are closer to each other (in the KS sense) than for the original data set ${\cal D}_J$. }
 \label{fig:gof-scheme}
\end{figure}

To evaluate the goodness of our fits, we again follow~\cite{Cla09}, to which we refer for an in-depth discussion of heavy-tailed distribution fitting. 
The whole goodness-of-fit estimation is summarized in Fig.~\ref{fig:gof-scheme}. 

Let us denote by $\theta_J$ the parameters of the distribution $P(X;\theta)$ (e.g., $\theta_J=\alpha$ for the power law distribution), fitted to the data set ${\cal D}_J$. 
We generate $5000$ sets of synthetic data $\tilde{\cal D}_i$, $i=1,...,5000$, each of them composed of $A_J^{\rm tot} = |{\cal D}_J|$ integer numbers, drawn randomly from the probability distribution $P_J = P(X;\theta_J)$. 
For each of these synthetic data sets $\tilde{\cal D}_i$, we perform again a MLE to fit the same distribution $P(X;\theta)$, yielding parameters $\tilde{\theta}_i$ and the distribution $P_i = P(X;\tilde{\theta}_i)$. 

The goodness-of-fit then relies on how well $F^{\rm e}$, the empirical cumulative distribution function (ECDF) for a given set of data, matches $F^{\rm t}$, the theoretical cumulative distribution function (TCDF) of its fitted distribution. 
We define
\begin{align}
 F^{\rm e}_i(k) &= \frac{\#\{n\in\tilde{\cal D}_i\colon n\leq k\}}{\#\tilde{\cal D}_i}\, , & F^{\rm t}_i(k) &= P(n\leq k;\theta_i)\, ,
\end{align}
and $F^{\rm e}_J$ and $F^{\rm t}_J$ are defined similarly with the data set ${\cal D}_J$. 

The $p$-value of the goodness-of-fit is then given by 
\begin{align}
 p &= \frac{\#\{i\colon d_{\rm KS}(F_i^{\rm e},F_i^{\rm t}) > d_{\rm KS}(F^{\rm e}_J,F^{\rm t}_J)\}}{5000}\, ,
\end{align}
where the \emph{Kolmogorov-Smirnov distance} between two cumulative distribution functions $F_1$ and $F_2$ is defined as the maximum difference between them, i.e.,
\begin{align}
 d_{\rm KS}(F_1,F_2) &= \max_k|F_1(k) - F_2(k)|\, .
\end{align}
Namely, $p$ is the proportion of synthetic data sets that are further from the theoretical distribution (in the Kolmogorov-Smirnov sense) than the analyzed data set. 
The fit is rejected if $p<5\%$, and considered as \emph{good} otherwise [see~\cite{Cla09} for more details]. 

This goodness-of-fit estimation is performed for each journal $J\in{\cal J}$ and each distribution listed above (power law, power law with cutoff, and Yule-Simon). 
The results are presented in Table~\ref{tab:fit_gof} and the resulting distributions together with the data are shown in Figs.~\ref{fig:prl_prd}, \ref{fig:full12}, and in the Supplementary Figure~\ref{fig:s1}. 

As can be seen in Figs.~\ref{fig:prl_prd}, \ref{fig:full12}, and Supplementary Figure~\ref{fig:s1}, the power law distribution is a poor fit for all data, its $p$-value being zero for all journals. 
Indeed, for most of the journals, the tail of the data set is lighter than the tail of its power law fit (black dashed lines). 
For three journals (namely SCI, PLC, CHA), the $p$-value of the power law with cutoff is larger than $5\%$ and it seems to be a rather good fit, and for two others (NEM and SIA), the Yule-Simon distribution cannot be excluded.

\section{General dynamics}
We argue that the heavy-tailedness observed in the previous section is likely to be a consequence of a \emph{preferential attachment} or \emph{cumulative advantage} process. 
Many social processes are ruled by the so called preferential attachment~[\cite{Jeo03}], also called cumulative advantage. 
Scientific co-authorship~[\cite{Bar02}], citations~[\cite{deS76,Eom11}], and performance of scientific institutions~[\cite{van07}] are apparently no exception to the rule. 
For instance, according to \cite{Eom11}, the probability that a paper will get a new citation at time $t$ is proportional to the number of citations this paper already has at time $t$. 

Such processes naturally lead to power laws in the relations between characteristics of the systems of interest. 
For instance, \cite{Kat99} showed that the number of citation a scientific community gets is a power law of the number of publications in this community, with positive exponent ($\approx 1.27$). 
More recently, \cite{Bet10} illustrate that the \emph{Gross Metropolitan Product} of a city is a power law of its population, with positive exponent ($\approx 1.126$). 
In a similar spirit, \cite{Bar99} showed that the empirical probability that a web page is targeted by $k$ other pages follows a power law with negative exponent ($\approx -2.1$). 

It is reasonable to expect that the evolution of the number of papers published by an author in a given journal is described by a similar preferential attachment process. 
We support the hypothesis of a preferential attachment or cumulative advantage process by two distinct but similar analysis of publication data. 

~

{\bf Remark.}
{\it Notice that even though we refer to the two analysis below as \emph{preferential attachment} and \emph{cumulative advantage} respectively, these two denominations fundamentally refer to the same general process~[\cite{Per14}]. 
The main reason for us to use these two denominations is to distinguish the two analysis. 
Furthermore, the line of reasoning underlying each of our analysis is inspired by the definition of the corresponding notion ("preferential attachment" or "cumulative advantage").  
}

\subsection{Preferential attachment}
Heuristically, our first argument is that if an author published a lot of papers in a journal, it means (i) that they write a lot of papers, and (ii) that their research topic is well-aligned with the scope of the journal (for specialized journals), or that the scientific impact of this author's research matches the standards of the journal (for interdisciplinary journals). 
Assumptions (i) and (ii) together imply that this author is likely to publish again in this journal. 
We refer to this process as \emph{preferential attachment}. 

\begin{figure*}
 \centering
 \includegraphics[width=\textwidth]{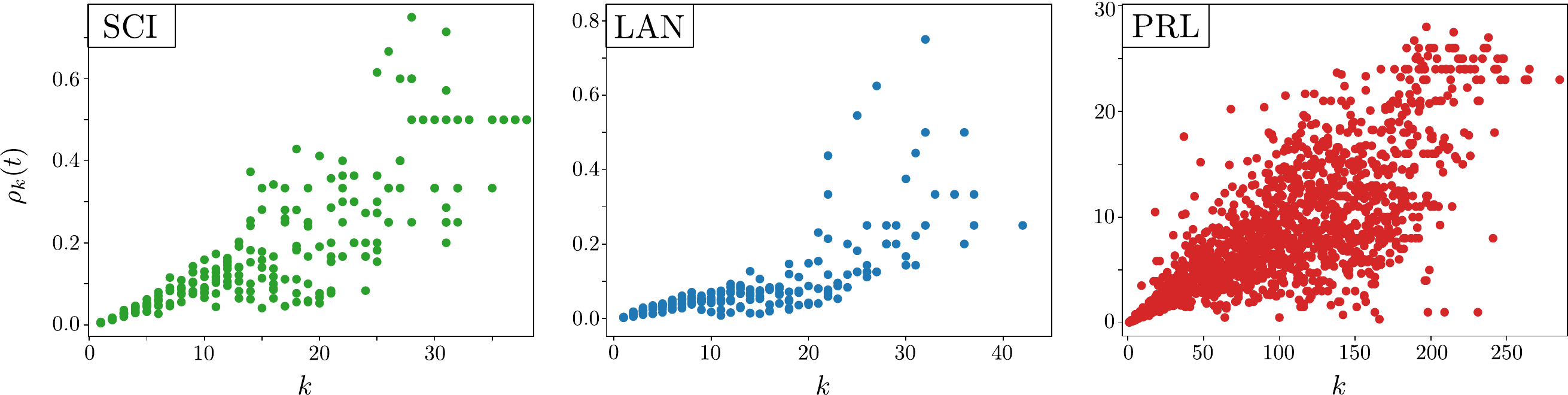}
 \caption{Average number of papers published within year $t\in\{1999,...,2008\}$, for authors in the set ${\cal S}(k,t)$, as a function of $k$, for SCI, LAN, and PRL. 
 Each points correspond to one of the years in  $\{1999,...,2008\}$ (hence multiple points for the same value of $k$). 
 The Pearson correlation coefficients of the point clouds are respectively $r_{\rm SCI}\approx 0.714$, $r_{\rm LAN}\approx0.707$, and $r_{\rm PRL}\approx0.763$, all larger than $0.7$, suggesting a relation close to linear. 
 For SCI (resp. LAN and PRL), $14$ points (resp. $12$ points and $2$ points) are left out of the frame, for sake of readability. }
 \label{fig:2}
\end{figure*}

The above heuristic can be made more rigorous. 
For a given journal and for $k,t\in\mathbb{Z}_{\geq 0}$, we define:
\begin{itemize}
 \item ${\cal S}(k,t)$: the set of all authors who have published $k$ papers on December 31st of year $t-1$;
 \item $A_k(t) = \#{\cal S}(k,t)$: the number of authors in the set ${\cal S}(k,t)$;
 \item $N_k(t)$: the number of papers published during year $t$ by all the authors in the set ${\cal S}(k,t)$;
 \item $\rho_k(t) = N_k(t)/A_k(t)\in\mathbb{R}$: the average number of papers published during year $t$, by the authors in the set ${\cal S}(k,t)$.
\end{itemize}
In Fig.~\ref{fig:2}, we plot the values of $\rho_k(t)$ with respect to the number of papers $k$ for years $t\in\{1999,...,2008\}$ for SCI, LAN, and PRL (each point corresponds to one year $t$ and one number of papers $k$). 
For each of the three journals, these values have a linear correlation coefficient larger than $0.7$, supporting a fairly good linear dependence, 
\begin{align}\label{eq:prop}
 \rho_k(t) &\sim k\, . 
\end{align}
Note that, for each year considered, we do not take into account authors who did not publish, because the majority of those are not active anymore. 

The empirical probability that a new paper is signed by an author with $k$ papers is then close to be proportional to $k$. 
\cite{Kra00} rigorously proved that, if the relation in Eq.~\eqref{eq:prop} was exactly proportional, then after a long enough time, the distribution of the number of papers over the set of authors would be a power law with exponent $\alpha\leq -2$. 
The fact that the relation~\eqref{eq:prop} is not exactly proportional, but close to be, probably explains that the observed distributions have tails that are heavy, but lighter than the power law, as suggested in Figs.~\ref{fig:prl_prd}, \ref{fig:full12}.

\subsection{Cumulative advantage}
The concept of \emph{cumulative advantage}, which is directly related to preferential attachment, has been derived from the seminal work of Merton~[\cite{Mer68,Mer88}] and Price [\cite{deS76}], and the follow-up by Katz~[\cite{Kat99}]. 
Cumulative advantage emphasizes that an initial advantage leads to a disproportionate advantage in the future. 
For instance, it has been shown that, if author $i$ has twice as many publications as author $j$, then they are likely to get more than twice as many citations~[\cite{Kat99}]. 

In the context of interest for this article, cumulative advantage translates as follows. 
Assume that author $i$ and author $j$ have respectively $n_i(t_0)$ and $n_j(t_0)$ papers in a journal at time $t_0$, with a ratio $\eta_{ij}(t_0) = n_i(t_0)/n_j(t_0) > 1$. 
Then cumulative advantage means that, at a later time $t_1>t_0$, the ratio $\eta_{ij}(t_1)\geq\eta_{ij}(t_0)$, implying that author $i$ gains a disproportional advantage over time. 
Mathematically speaking, cumulative advantage implies the following equivalences, 
\begin{align}\label{eq:incr}
 n_i(t_0) &\geq n_j(t_0) &&\iff & \frac{n_i(t_0)}{n_j(t_0)} &\leq \frac{n_i(t_1)}{n_j(t_1)} &&\iff & \frac{n_i(t_1)}{n_i(t_0)} &\geq \frac{n_j(t_1)}{n_j(t_0)} &&\iff & \xi_i(t_0,t_1) &\geq \xi_j(t_0,t_1)\, ,
\end{align}
where we defined $\xi_i(t,s) = n_i(s)/n_i(t)$, and where equalities hold if the relation in Eq.~\eqref{eq:prop} is exact. 

\begin{figure*}
 \centering
 \includegraphics[width=\textwidth]{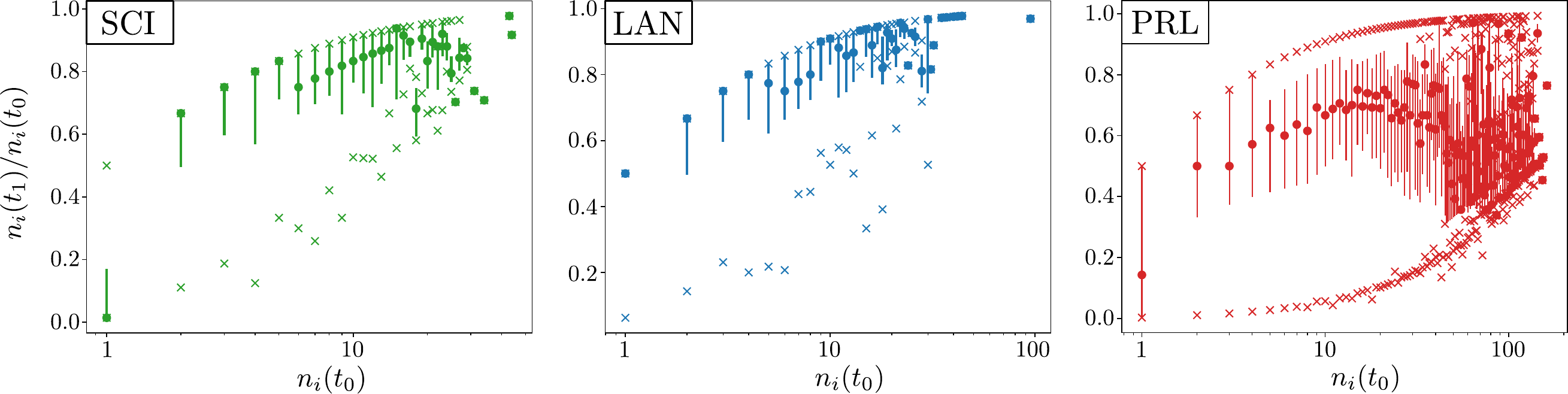}
 \caption{Statistics of the ratio $\xi_i$ between the number of papers in 1999 and in 2008 as a function of the number $n_i$ of papers in 1999, in the three journals SCI (left), LAN (center), and PRL (right). 
 For each value of $n_i(1999)$, there are multiple authors with this number of papers in 1999. 
 Among these authors, the dots show the median value of $\xi_i$, the bar covers the second and third quartiles, and the crosses are the maximal and minimal values. 
 Whereas there is not an exact increase of the values, there is anyway an increasing trend of $\xi_i$ with respect to $n_i$, supporting the presence of a cumulative advantage process. }
 \label{fig:2bis}
\end{figure*}

In order to support the presence of a cumulative advantage in the publication within the journals SCI, LAN, and PRL, we computed $\xi_i(1999,2008)$ for each author who published between 1999 and 2008. 
The statistics of $\xi_i$ are shown in Fig.~\ref{fig:2bis} as a function of the initial number of papers $n_i(1999)$. 
Even though the data are not perfectly conclusive, we clearly observe an increasing trend of $\xi_i$ as a function of $n_i$, suggesting that the relation of Eq.~\eqref{eq:incr} may be satisfied. 
This observation supports (at least partly) a cumulative advantage process, and henceforth the presence of a power law. 

The increasing trends in Fig.~\ref{fig:2bis} even suggest a superlinear cumulative advantage [\cite{Zho07,Kra08}]. 
Indeed, as mentioned above, if the relation Eq.~\eqref{eq:prop} was exact, $\xi_i(t_0,t_1)$ would be constant with respect to $n_i(t_0)$. 
In such a case, the heavy-tailed distribution observed in Figs.~\ref{fig:prl_prd}, \ref{fig:full12}, and \ref{fig:s1} would be the transient state of the distribution discussed by \cite{Kra08}. 
A more in-depth analysis of the possibility of a superlinear cumulative advantage could be done, following the calibration approach proposed by \cite{Zad15}, but goes beyond the purpose of this article and will be treated in future work.

\section{Key players}\label{sec:kplayers}
\begin{figure*}
 \centering
 \includegraphics[width=.95\textwidth]{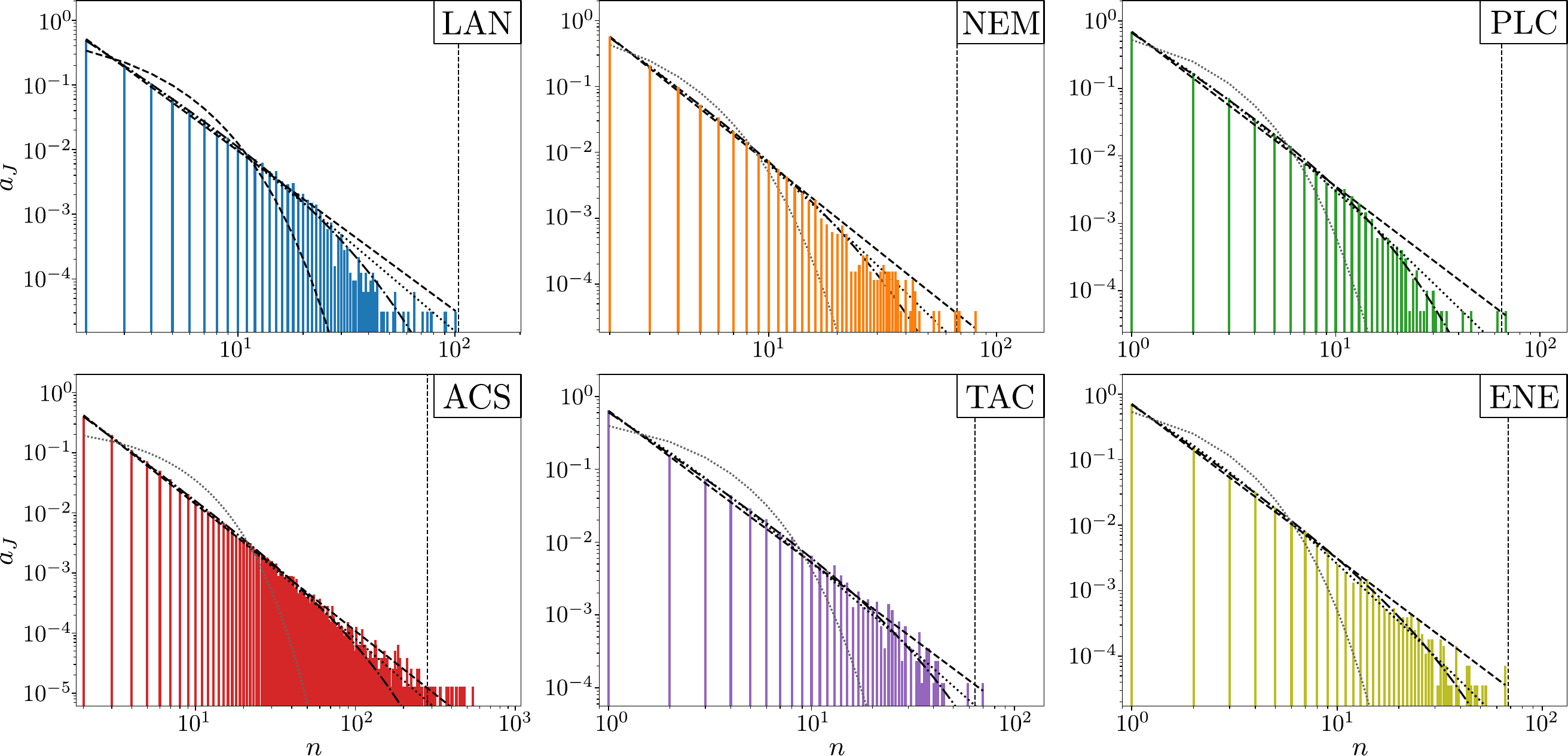}
 \caption{Supplementary Figure. 
 Histograms of the number of papers $n$ published in the six journals indicated in the insets, among the authors who published in these journals (see Table~\ref{tab:journals} for legends). 
 As in Figs.~\ref{fig:prl_prd} and \ref{fig:full12}, for each value of $n$, the height of the bar gives the proportion of authors who published $n$ articles in the corresponding journal. 
 The gray dotted line is the exponential fit of the data, emphasizing that the distribution is heavy-tailed. 
 We show the best fit for a power law distribution (dashed black), power law with cutoff (dash-dotted black), and Yule-Simon distribution (dotted black). 
 The vertical dashed line indicates the theoretical maximal number of published papers if the distribution was the fitted power law. }
 \label{fig:s1}
\end{figure*}

The general distribution of the number of papers per author is quite clear in our analysis, it seems to be somewhere between an exponential distribution and a power law. 
The power law having the heaviest tail of the three distributions considered (power law, power law with cutoff, and Yule-Simon), we use it to estimate an upper bound on the number of papers published by an author for each journal.
Assuming that the data are well-described by the power law distribution in Eq.~\eqref{eq:pl}, one can compute the number of authors with $n$ papers in journal $J$, $A_n \approx A_J^{\rm tot} C_\alpha n^{-\alpha}$. 
Setting this number to $A_n=1$, the maximal number of papers is given by $n_{\max} \approx (A_J^{\rm tot} C_\alpha)^{\frac{1}{\alpha}}$, determining a theoretical upper bound on the number of papers published by an author for each journal, shown as the vertical dashed lines in Figs.~\ref{fig:prl_prd}, \ref{fig:full12}, and \ref{fig:s1}. 
 
In some journals (see e.g., PNA, CHA, SIA, and AMA in Fig.~\ref{fig:full12}, and  NEM and ACS in the Supplementary Figure~\ref{fig:s1}), it appears that, some authors, which we refer to as \emph{key players}, publish significantly more papers in a journal than what the power law would predict.
Note that we checked that these key players are not artifacts due to multiple authors having the same name which would count as the same person. 

In order to make the data of different journals more comparable, we restricted our investigation to the early years between 1900 (earliest possible in WoS) and the year in parenthesis in the second column of Table~\ref{tab:journals} for our first nine journals in the table. 
This yields a number of authors comparable to the three following journals in Table~\ref{tab:journals} (CHA, SIA, and AMA). 
The reduced number of authors is given in parenthesis in the third column of Table~\ref{tab:journals}. 
The resulting distributions are depicted in Fig.~\ref{fig:full_red9} and in the Supplementary Figure~\ref{fig:s2}, and the fitted parameters are detailed in Table~\ref{tab:fit_gof_red}. 
It appears from Fig.~\ref{fig:full_red9} and the Supplementary Figure~\ref{fig:s2} that for such reduced number of authors, the overshoot of some authors is more systematic, suggesting that in the early years of scientific journals, there is usually a few very prolific authors publishing in it at a rather high rate. 

\begin{figure*}
 \centering
 \includegraphics[width=\textwidth]{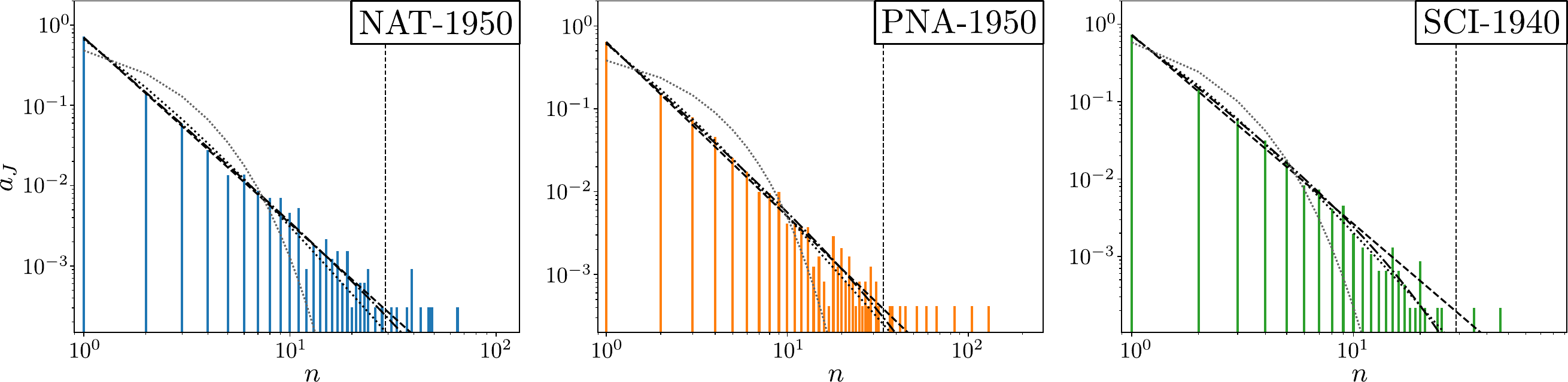}
 \caption{Histograms of the number of papers $n$ published in the six journals indicated in the insets, among the authors who published in these journals (see Table~\ref{tab:journals} for legends). 
 Data are restricted to the years between 1900 (earliest possible in WoS) and the years indicated in the insets. 
 The number of authors covered is given in parenthesis in the third column of Table~\ref{tab:journals}. . 
 As in Fig.~\ref{fig:prl_prd} and \ref{fig:full12}, for each value of $n$, the height of the bar gives the proportion of authors who published $n$ articles in the corresponding journal. 
 We show the best fit for a power law distribution (dashed black), power law with cutoff (dash-dotted black), and Yule-Simon distribution (dotted black). 
 The vertical dashed line indicates the theoretical maximal number of published papers if the distribution was the fitted power law (see Sec.~\ref{sec:kplayers}). 
 We observe an almost systematic exceeding of the number of papers published by some authors. 
 The same plot for other journals is available in the Supplementary Figure~\ref{fig:s2}.}
 \label{fig:full_red9}
\end{figure*}

\begin{table}
 \centering
 \begin{tabular}{l||c|c||c|c|c||c|c}
  & \multicolumn{2}{c||}{PL} & \multicolumn{3}{c||}{PLwC} & \multicolumn{2}{c}{Y-S} \\
  & $\alpha$ & $p$ [\%] & $\beta$ & $\gamma$ & $p$ [\%] & $\rho$ & $p$ [\%] \\
  \hline
  \hline NAT & $\bf 2.32$ & $\bf 29.4$ & $\bf 2.23$ & $\bf 0.016$ & $\bf 6.0$ & $2.98$ & $0.0$ \\
  \hline PNA & $2.10$ & $0.1$ & $\bf 1.96$ & $\bf 0.02$ & $\bf 15.0$ & $\bf 2.55$ & $\bf 6.3$ \\
  \hline SCI & $2.44$ & $0.0$ & $\bf 2.13$ & $\bf 0.09$ & $\bf 72.0$ & $3.37$ & $4.7$ \\
  \hline LAN & $2.25$ & $0.0$ & $\bf 1.81$ & $\bf 0.11$ & $\bf 30.2$ & $2.91$ & $2.5$ \\
  \hline NEM & $2.27$ & $0.9$ & $2.06$ & $0.04$ & $4.4$ & $2.91$ & $0.0$ \\
  \hline PLC & $2.59$ & $0.0$ & $2.12$ & $0.16$ & $0.3$ & $\bf 3.82$ & $\bf 54.7$ \\
  \hline ACS & $2.06$ & $0.0$ & $1.89$ & $0.02$ & $0.1$ & $\bf 2.46$ & $\bf 64.0$ \\
  \hline TAC & $2.32$ & $0.0$ & $\bf 2.06$ & $\bf 0.06$ & $\bf 23.7$ & $3.04$ & $0.1$ \\
  \hline ENE & $2.69$ & $0.8$ & $\bf 2.50$ & $\bf 0.06$ & $\bf 94.5$ & $4.06$ & $0.0$  
 \end{tabular}
 \caption{Fitted parameters and $p$-value of the goodness-of-fit for power law (PL), power law with cutoff (PLwC), and Yule-Simon (Y-S) distributions, for the 9 journals with reduced time span. 
 We see that the only data that are well-approximated by the power law are for NAT when reduced to the first 3374 entries of WoS. 
 The power law with cutoff, however, seems to be a good fit for the reduced data of six journals (NAT, PNA, SCI, LAN, TAC, and ENE). 
 ENE is particularly well-fitted by the power law with cutoff. 
 Finally, the Yule-Simon distribution seems to correctly fit the distribution of PAN, PLC, and ACS. 
 For the other journals, none of the distributions seem to fit the data appropriately. 
 Remark that the reduced data of NAT and PNA are correctly fitted for two distributions indicating that the amount of data is probably not sufficient for a good fit. }
 \label{tab:fit_gof_red}
\end{table}

Considering the results of the fitting, in Table~\ref{tab:fit_gof_red}, we observe better agreements than for the full data sets. 
This probably indicates that the sample size is not large enough to accurately fit heavy-tailed distributions, which obviously need large samples. 
The fact that NAT and PNA are well-fitted by two distributions, also indicates that the reduced data sets are not large enough to be conclusive.

\section{Modeling}
We observe in Figs.~\ref{fig:prl_prd}, \ref{fig:full12}, and in the Supplementary Figure~\ref{fig:s1} that for old journals where a lot of papers are published, the tail of the histogram has a rather fast decay after a heavy-tailed regime (this is particularly striking in PRL and PRD, Fig.~\ref{fig:prl_prd}). 
We explain this observation by the fact that the number of publications of a given author depends on two parameters, namely their publication rate and the length of their career. 
Both these quantities are bounded in practice and even if it is possible to publish a very large number of papers in a given journal, there is a practical limit to this number. 
We hypothesize that the decay in the histograms of long-living journals comes from the finiteness of publication rates and career lengths. 

To support our hypothesis, we propose a model to generate data sets that mimic the distributions observed above. 
As discussed, this model is built on two main dynamics. 
Fundamentally, it is a \emph{preferential attachment} process, where the likelihood that a researcher is in the author's list of a new paper is proportional to the number of papers this researcher already has in this journal. 
But in addition, it is refined with a \emph{limited career span}, requiring that after some time, the likelihood that a researcher publishes a new paper decreases to reach zero after they retire. 

\begin{figure*}
 \centering
 \includegraphics[width=.55\textwidth]{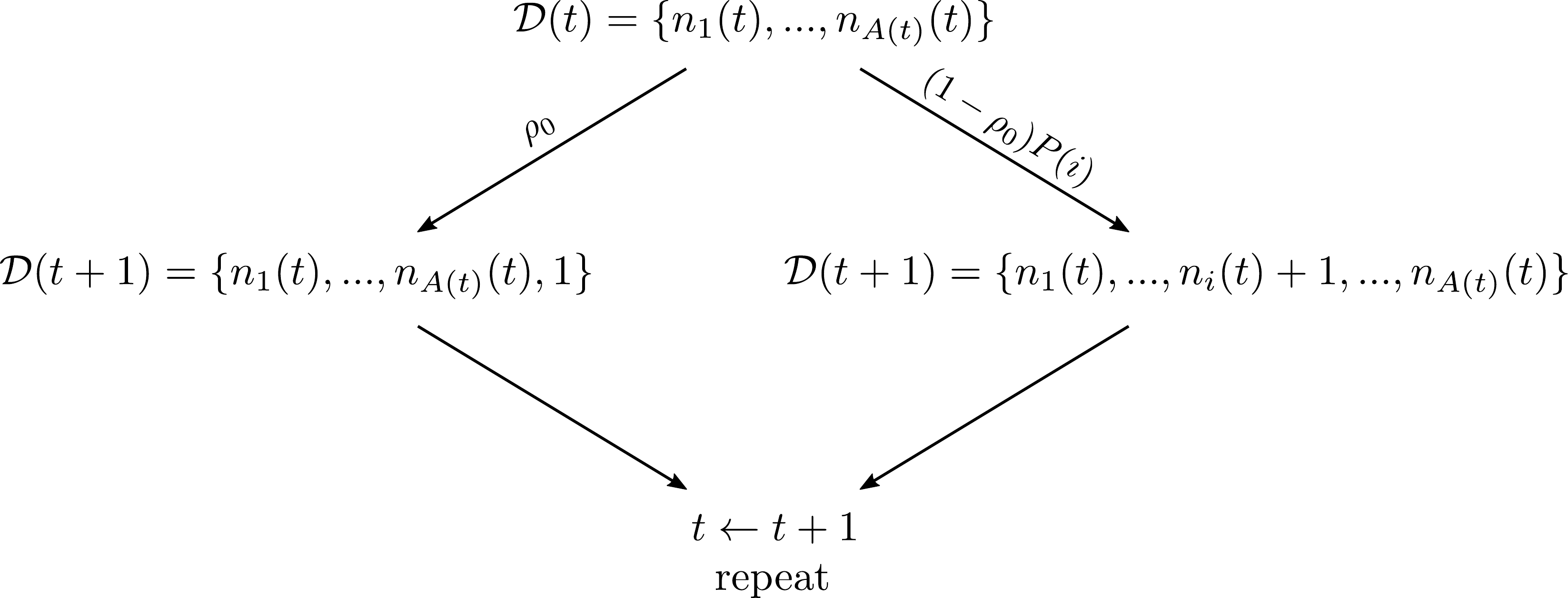}
 \qquad
 \includegraphics[width=.35\textwidth]{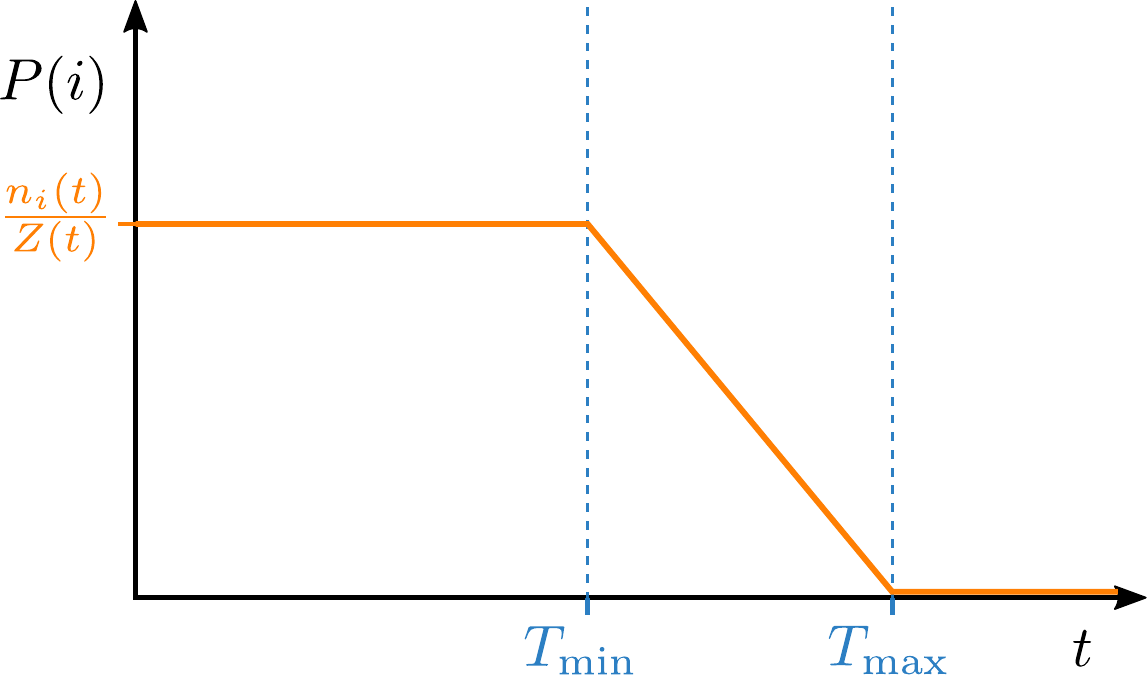}
 \caption{Left: Scheme of the iterative process generating the synthetic distribution of number of publication per author in a journal. 
 Right: Illustration of the probability that a new paper is attributed to author $i$, knowing that they have already published in the past.}
 \label{fig:model-scheme}
\end{figure*}

The models is based on five parameters:
\begin{itemize}
 \item $N_{\rm y}\in\mathbb{Z}_{\geq 0}$: The number of years, i.e., number of iteration, over which the model is run. 
 \item $N_{\rm p}\in\mathbb{Z}_{\geq 0}$: The number of papers that are published every year in the synthetic journal;
 \item $\rho_0\in[0,1]$: The proportion of papers that are authored by new researchers who have not yet published in the synthetic journal;
 \item $T_{\min}, T_{\max}\in\mathbb{Z}_{\geq 0}$: The likelihood that an author publishes a new paper decreases linearly after their $T_{\min}$th year of activity, until reaching zero at their $T_{\max}$th year of activity. 
 We illustrate this likelihood in Fig.~\ref{fig:model-scheme}
\end{itemize}

The model is arbitrarily initialized with some number of authors each with a few papers in the synthetic journal, gathered in the data set ${\cal D}(0) = \{n_1(0), n_2(0),...,n_{A(0)}(0)\}$. 
Then for each year $t\in\{1,...,N_{\rm y}\}$ where the model is run, $N_{\rm p}$ papers are attributed randomly either to new authors (i.e., who have not yet published) with probability $\rho_0$, or to an existing author with probability $1-\rho_0$. 
If it is attributed to an existing author, the probability that it is attributed to author $i$ is:
\begin{itemize}
 \item proportional to $n_i(t)$, the number of papers published by $i$ at year $t$; 
 \item linearly decreasing for $T_i(t)\in[T_{\min},T_{\max}]$, where $T_i(t)$ is the \emph{"academic age"} of $i$, which is the number of iteration between $t$ and the first publication year of $i$.
\end{itemize} 
Mathematically, knowing that the new paper is attributed to an existing author, the probability that it is attributed to author $i$ at year $t$ is given by 
\begin{align}
 P(i) &= \frac{1}{Z(t)} n_i(t) \min\left\{1,\frac{T_{\max} - T_i(t)}{T_{\max} - T_{\min}} \right\}\, ,
\end{align}
where $Z(y)$ is the appropriate normalizing factor. 
The actual implementation of this model is available online~[\cite{repo}]. 

\begin{figure*}
 \centering
 \includegraphics[width=\textwidth]{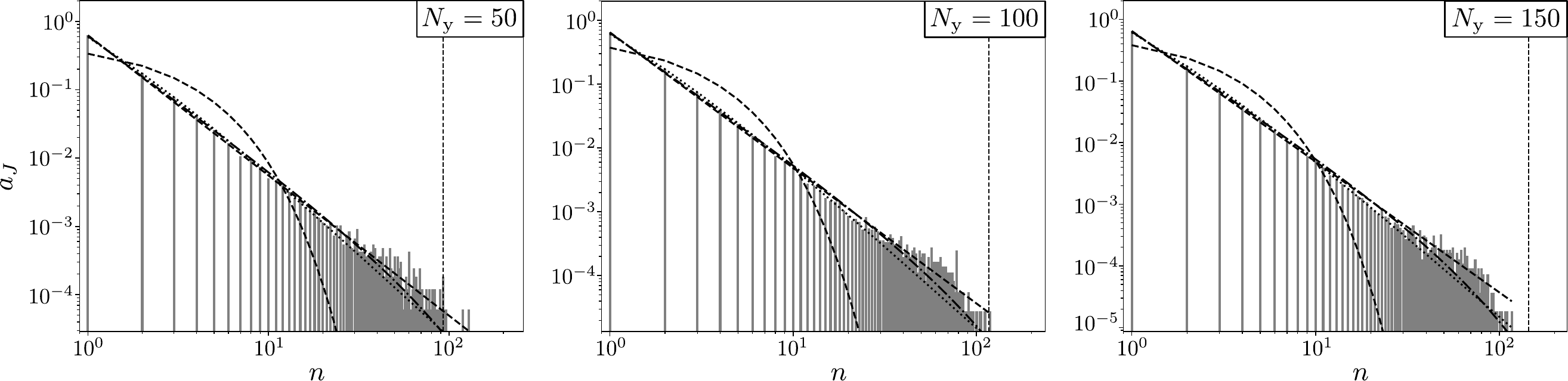}
 \caption{Histograms of the outcome of our synthetic data generator, for different value of the journal life spa $N_{\rm y}$. 
 Fixed parameters are $N_{\rm p} = 1000$, $\rho_0 = 0.5$, $T_{\min} = 20$, $T_{\max} = 60$. 
 There is a clear similarity between the shapes of these synthetic distributions and those of the actual data. }
 \label{fig:synth}
\end{figure*}

Histograms of the outcome of this model are illustrated in Fig.~\ref{fig:synth} and the fitted parameters are in Table~\ref{tab:fit_gof_synth}. 
We observe a clear similarity between the histograms for synthetic and real data. 
Namely, for short lifetime ($N_{\rm y}=50$), some authors beat the power law and exceed the number of papers that would be expected, as is observed in Fig.~\ref{fig:full12} for CHA, SIA, and AMA. 
For longer lifetime ($N_{\rm y}=150$) the tail of the distribution decays and loses its heaviness similarly as PRL and PRD in Fig.~\ref{fig:prl_prd}. 

\begin{table}
 \centering
 \begin{tabular}{l||c|c||c|c|c||c|c}
  & \multicolumn{2}{c||}{PL} & \multicolumn{3}{c||}{PLwC} & \multicolumn{2}{c}{Y-S} \\
  & $\alpha$ & $p$ [\%] & $\beta$ & $\gamma$ & $p$ [\%] & $\rho$ & $p$ [\%] \\
  \hline
  \hline $N_{\rm y}=50$ & $2.05$ & $0.0$ & $1.94$ & $0.013$ & $0.0$ & $2.44$ & $0.2$ \\
  \hline $N_{\rm y}=100$ & $2.12$ & $0.0$ & $2.03$ & $0.01$ & $0.0$ & $2.58$ & $0.0$ \\
  \hline $N_{\rm y}=150$ & $2.12$ & $0.0$ & $2.01$ & $0.02$ & $0.0$ & $2.58$ & $0.06$  
 \end{tabular}
 \caption{Fitted parameters and $p$-value of the goodness-of-fit for power law (PL) and power law with cutoff (PLwC), and Yule-Simon (Y-S) distributions on the synthetic histograms of Fig.~\ref{fig:synth}. 
 None of the goodness-of-fit test is conclusive, but the values of the fitted parameters are very similar to what is observed in actual data. 
 }
 \label{tab:fit_gof_synth}
\end{table}

These observations advocate in favor of the hypothesis that the two main ingredient in the description of the evolution of the authorship within journals are both the \emph{preferential attachment} and the \emph{finiteness of careers}.

\section{Discussion}
The main observation of our article is the heavy-tailed shape of the distribution of papers, which we explain by a preferential attachment or cumulative advantage process. 
Heavy-tailedness in distributions related to scientific publications, especially in citation or collaboration networks, has widely been documented~[\cite{deS76,Eom11}]. 
We showed that heavy-tailedness is preserved when restricting the analysis to a single journal. 

Interestingly, our analysis suggests that the distribution does not follow a power law, but has a slightly lighter tail. 
Whereas we have not been able to unequivocally identify a canonical distribution, we demonstrated that a power law with cutoff or a Yule-Simon distribution seem to be better fits to the data than the power law. 

We argue that the observed heavy-tailedness of the distribution follows from a preferential attachment process through three pieces of evidence. 
First, we showed that the probability that an author gets a new paper in a given journal at time $t$ is approximately proportional to the number of paper they already have in the very same journal. 
According to \cite{Kra00}, exact proportionality would lead to a power law. 
Therefore, it is likely that an approximate proportionality leads to an heavy-tailed distribution. 

Second, we emphasized an approximate cumulative advantage process, which also leads to power law behaviors. 
Whereas both what we refer to as preferential attachment and cumulative advantage are closely related, they display two underlying mechanisms explaining the heavy-tailedness of the distributions. 

Finally, we provided a mathematical model for generating synthetic data of number of papers in a given journal, where preferential attachment plays a crucial role. 
The similarity between the obtained distribution and the observed distributions also supports the claim of the heavy tails being driven by preferential attachment. 

Even though there seems to be a pattern in the data analyzed in this article, standard distributions (e.g., power law with cutoff, Yule-Simon) do not perfectly fit the data. 
More advanced fitting techniques could identify a common distribution for all journals, provided that one exists. 
A more refined explanation of the approximate preferential attachment taking place in scientific publishing could unravel with more certainty the source of the distributions observed in this article. 
Even though the preferential attachment has been emphasized in the past, the underlying reasons of this bias are intricate. 
Disentangling the impact of scientific factors (quality and novelty of the research) and more social ones (rank and reputation of the authors) in the publication process will be a key step towards a fair and square evaluation of scientists and their work.

\section*{Authors contributions} 
{\bf Robin Delabays:} Conceptualization, Data curation, Formal analysis, Investigation, Methodology, Software, Validation, Visualization, Writing -- original draft, Writing -- review \& editing.
{\bf Melvyn Tyloo:} Conceptualization, Methodology, Writing -- review \& editing.

\section*{Funding information}
Both authors were partly supported by the Swiss National Science Foundation under grant number 200020\_182050.
RD was supported by the Swiss National Science Foundation under grant number P400P2\_194359. 

\section*{Competing interests} 
The authors declare no competing interest. 

\section*{Data availability} 
The data were extracted from \url{www.webofscience.com} and we cannot share it openly. 
The code for synthetic data generation is available online~[\cite{repo}]. 

\vspace{5mm}

\hrule

\newpage

\begin{figure*}
 \centering
 \includegraphics[width=.95\textwidth]{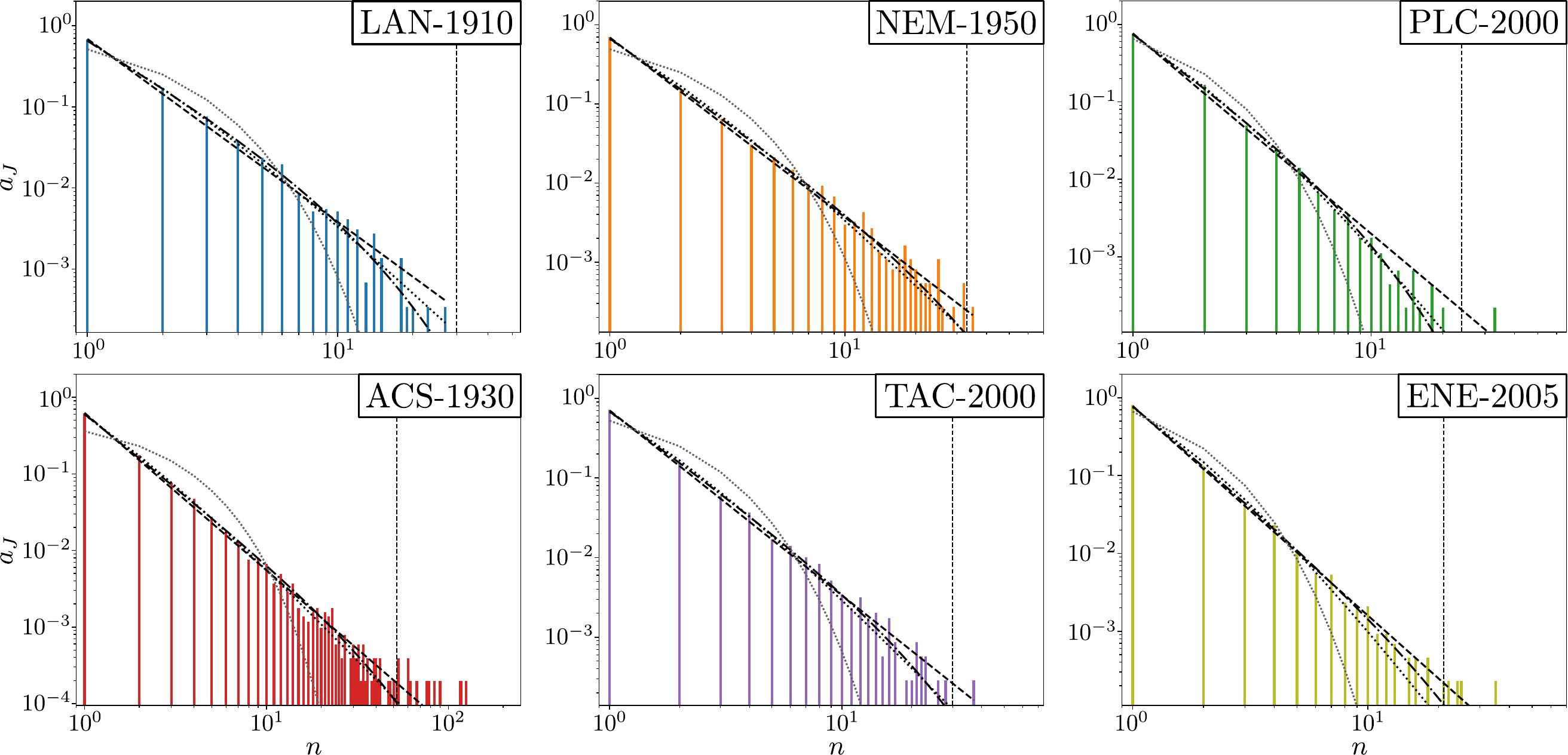}
 \caption{Supplementary Figure. 
Histograms of the number of papers $n$ published in the six journals indicated in the insets, among the authors who published in these journals (see Table~\ref{tab:journals} for legends). 
 Data are restricted to the years between 1900 (earliest possible in WoS) and the years indicated in the insets. 
 The number of authors covered is given in parenthesis in the third column of Table~\ref{tab:journals}. . 
 As in Fig.~\ref{fig:prl_prd} and \ref{fig:full12}, for each value of $n$, the height of the bar gives the proportion of authors who published $n$ articles in the corresponding journal. We show the best fit for a power law distribution (dashed black), power law with cutoff (dash-dotted black), and Yule-Simon distribution (dotted black). 
 The vertical dashed line indicates the theoretical maximal number of published papers if the distribution was the fitted power law. 
 We observe an almost systematic exceeding of the number of papers published by some authors. }
 \label{fig:s2}
\end{figure*}

\end{document}